\documentclass[conference]{IEEEtran}

\usepackage{graphicx} 
\usepackage{cite}
\usepackage{amsmath,amssymb,amsfonts}

\usepackage{graphicx}
\usepackage{textcomp}
\usepackage{xcolor}
\usepackage{url}
\usepackage{subcaption}
\usepackage{algorithm}
\usepackage{algpseudocode}
\usepackage{geometry}
\usepackage{tikz}

\newcommand\copyrighttext{%
  \footnotesize 2026 IEEE. Personal use of this material is permitted. Permission from IEEE must be obtained for all other uses, in any current or future media, including reprinting/republishing this material for advertising or promotional purposes, creating new collective works, for resale or redistribution to servers or lists, or reuse of any copyrighted component of this work in other works.}
\newcommand\copyrightnotice{%
\begin{tikzpicture}[remember picture,overlay]
\node[anchor=south,yshift=10pt] at (current page.south) {\fbox{\parbox{\dimexpr\textwidth-\fboxsep-\fboxrule\relax}{\copyrighttext}}};
\end{tikzpicture}%
}

\begin{document}


\title{\fontsize{18.9}{18.9}\selectfont Low-Complexity, Space Splitting-based User Selection in MU-MIMO for Massive Connectivity and AI-Native Traffic}
\author{\IEEEauthorblockN{João Paulo S. H. Lima, Marcin L. Filo, Chathura Jayawardena, and Konstantinos Nikitopoulos 
}
\IEEEauthorblockA{Wireless Systems Lab, 5G and 6G Innovation Centre, University of Surrey, Guildford, GU2 7XH, UK \\ E-mails: \{j.lima, m.filo, m.jayawardena, k.nikitopoulos\}@surrey.ac.uk
}
\vspace{-31pt}
}

\maketitle
\copyrightnotice

\begin{abstract}
The rise of Artificial Intelligence (AI)-driven services, machine-type communications, and massive Internet of Things (IoT) deployments is reshaping wireless traffic toward dense, uplink-oriented, bursty, and latency-critical patterns.
In these regimes, Multi-User Multiple-Input Multiple-Output (MU-MIMO) is essential to support massive concurrent connectivity through spatial multiplexing. However, the need for frequent, low-latency scheduling decisions exposes fundamental scalability barriers in existing user selection approaches. The inherently combinatorial nature of MU-MIMO user selection leads computational complexity to grow rapidly with both the number of candidate users and spatial layers, rendering existing near-optimal heuristic methods impractical in dense and highly dynamic scenarios.
This paper introduces the Space Splitting-based User Selection (SS-US) 
algorithm, a complexity barrier-breaking, massively parallelizable method that departs from subset-based selection by constructing orthonormal spatial bases and independently matching users to spatial directions. Simulation results across diverse MIMO configurations, channel conditions, and user densities show that SS-US reduces computational complexity by over three orders of magnitude while achieving spectral efficiency comparable to state-of-the-art practical baselines.
\end{abstract}

\begin{IEEEkeywords}
MU-MIMO, User Selection, Space Splitting, Complexity
\vspace{-5pt}
\end{IEEEkeywords}
\vspace{-8pt}
\section{Introduction}
\label{sec:intro}

The nature of wireless traffic patterns is rapidly evolving.
Artificial Intelligence (AI)-enabled applications and agents, together with the proliferation of sensors and machine-type communications, are fundamentally reshaping network connectivity patterns.
Unlike traditional broadband services, where the uplink typically represents only about 10\% of total traffic~\cite{ericssonmobrep}, AI-native and machine-driven communications generate highly diverse traffic profiles, characterized by burstiness, heterogeneity, and a strong shift toward uplink-oriented transmissions.
Recent reports indicate that the uplink share may approach 50\% in specific scenarios~\cite{nokiareport,ericssonmobrep}.
Beyond the high volume, these transmissions are often more latency- and reliability-critical, as required by interactive or multi-stage AI workloads and applications such as extended and virtual reality (XR/VR), Industry~5.0, and vehicular intelligence in smart cities~\cite{xr-mag,xiang2023advanced,karagiannis2011vehicular}.
Projections further suggest that the number of Internet of Things (IoT) devices will exceed 40~billion by 2030~\cite{iot-analytics}, while AI-related traffic is expected to surpass 1000~EB (exabytes, or $10^{18}$ bytes!) per month by 2033~\cite{nokiareport}.
This combination of massive device density, highly variable activity patterns, and uplink-heavy traffic introduces new operational challenges for networks.
In particular, frequent scheduling decisions must be made under stringent latency constraints, requiring algorithmic solutions that are not only spectrally efficient but also highly scalable with low complexity.

In this context, Multiple-Input, Multiple-Output (MIMO), and especially Multi-User (MU) MIMO, remains a key enabler for supporting massive and simultaneous connectivity.
By exploiting spatial multiplexing, MU-MIMO allows multiple users to transmit concurrently, significantly improving spectrum utilization and system throughput.
However, emerging traffic scenarios with numerous candidate devices drive the need for higher MIMO orders, rendering the user selection process, which determines how and when users share spatial resources, a critical bottleneck.
For instance, selecting 4 out of 50 users for allocation results in $2.3\times 10^{5}$ combinations, whereas selecting 8 out of 100 users for allocation has $1.86\times 10^{11}$ possible subsets.
Traditional approaches~\cite{sched-surv-marcin,ket-sched-surv,sus} were largely designed for earlier traffic regimes and moderate MIMO dimensions.
As a result, they do not scale well with growing user densities and antenna counts, nor are they inherently designed to parallel execution, making it difficult to meet the tight latency and reconfigurability requirements of emerging scenarios.
At the same time, offloading user selection to hardware to address this challenge risks limiting flexibility and upgradability, as scheduling policies, in which user selection plays a central role, must be frequently adapted to evolving traffic scenarios.


Several heuristic approaches have been proposed and categorized in the literature based on their user selection strategy~\cite{sched-surv-marcin}.
Null-space projection methods, such as the Semi-orthogonal User Selection (SUS)~\cite{sus}, prioritize spatial orthogonality by comparing channel projections against a correlation threshold.
While computationally feasible, SUS relies on sequential null-space updates that limit parallel execution, and its repeated projection operations scale quadratically with the number of active MIMO layers (as analyzed in Sec.~\ref{subsec:eval-complexity}), challenging real-time operation in large-scale arrays.
Alternatively, greedy approaches, such as~\cite{gzf,mu-rme}, adopt a seed-dependent expansion that recursively estimates incremental gains under a chosen policy (e.g., sum-rate maximization). 
These methods often achieve near-optimum performance, at the cost of repeated heavy matrix operations (e.g., inversions), whose complexity grows prohibitively with the user pool size and MIMO order.
Subspace compatibility methods employ metrics such as chordal distance, used in the mCore+ scheduler~\cite{mcore-sched}, to assess spatial compatibility between users.
While their pairwise formulation enables partial parallelization of distance computations, enabling, for example, the scheduling of 100 users with four spatial layers within 500~$\mathrm{\mu s}$ in~\cite{mcore-sched}, this method relies on exhaustive evaluations of achievable rates across candidate combinations, leading to prohibitive complexity as the number of layers increases.
Overall, although these strategies strike different balances between computational complexity and scheduling gains, they often depend on inherently sequential selection steps that hinder parallelization or involve heavy combinatorial computations whose complexity grows rapidly with the number of candidate users or MIMO layers.
As a result, they are not optimized for the highly dense, latency-sensitive uplink scenarios imposed by the emerging traffic of AI agents and applications.

In this context, this work proposes, for the first time, the Space Splitting-based User Selection (SS-US), 
a novel low-complexity method that constructs nearly orthonormal user subsets to maximize spatial separability with minimal computation.
Unlike existing MU-MIMO user selection techniques, which rely on iterative or combinatorial exploration of user subsets, SS-US reformulates user selection as a directional matching problem in the spatial domain.
By partitioning the spatial degrees of freedom into orthonormal subspaces and independently selecting users that align closely with the basis vectors of each subspace, SS-US avoids repeated subset evaluation, matrix inversion, and iterative orthogonality refinement.
This design enables intrinsically parallel execution, bounded per-slot complexity, and explicit control over the performance–complexity balance through a small set of tunable parameters.
Simulation results across multiple MIMO configurations, channel conditions, and user densities show that SS-US reduces computational complexity by over three orders of magnitude (more than $2500\times$) while maintaining spectral efficiency comparable to practical, high-performing baselines.
These properties make SS-US a strong candidate for next-generation MU-MIMO systems operating in dense, uplink-oriented, and latency-sensitive environments driven by AI-native and machine-type communications.

The remainder of the paper is organized as follows: Section~\ref{sec:system-model} presents the system model, Section~\ref{sec:user-selection} introduces SS-US, and Section~\ref{sec:evaluation} reports the evaluation methodology, simulation results, and complexity analysis. Section~\ref{sec:conclusions} concludes the paper.

\vspace{-8pt}
\section{System Model}
\label{sec:system-model}

Consider a single-cell uplink scenario comprising a base station (BS) supporting linear MIMO signal processing, equipped with $M$ antennas and a set of $U$ user equipment devices (UEs), where $U \gg M$ represents a dense connectivity condition.
Each UE may have multiple antennas, but for simplicity, we assume that each UE transmits a single stream to the BS. Although the focus of this work is on the uplink, the proposed framework can be readily extended to the downlink.

The transmission is organized in Orthogonal Frequency-Division Multiplexing (OFDM) slots, defined by a grid of $N_{SC}$ subcarriers and $N_{OS}$ OFDM symbols (typically $N_{OS}=14$).
In compliance with 5G NR specifications, each resource block (RB) consists of 12 subcarriers, and the time-frequency grid is divided into resource elements (REs), where each RE spans one OFDM symbol by one subcarrier.
Resource block groups (RBGs) represent the minimum scheduling allocation unit, each composed of one or more RBs, yielding $B$ RBGs per slot.

\begin{figure}
    \centering
    \includegraphics[width=0.6\linewidth]{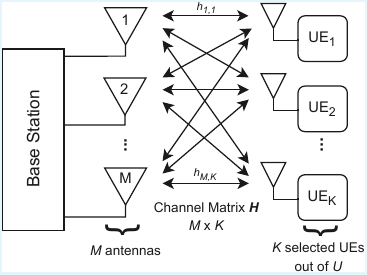}
    \caption{System model showing the channel matrix $H$, when $K$ single-antenna users are selected out of $U$ candidates.}
    \label{fig:sys-model}
    \vspace{-15pt}
\end{figure}

Let $\mathbf{H}_b \in \mathbb{C}^{M\times U}$ denote the channel matrix in RBG $b \in \left[ 1,B\right]$, where each column $\mathbf{h}_u \in \mathbb{C}^{M\times 1}$ represents the channel vector of user $u$, with $\lVert \mathbf{h}_{u} \rVert$ denoting its norm.
For each RBG, a subset of UEs $\mathcal{K}_{b} \subseteq \mathcal{U}$ is scheduled for simultaneous MU-MIMO transmission, as illustrated in Fig.~\ref{fig:sys-model}.
The number of selected UEs in RBG $b$ is $K_{b} = \left|\mathcal{K}_{b}\right| \leq M$, as we consider a linear MIMO processing at the BS (cases where $K_{b} = \left|\mathcal{K}_{b}\right| > M$ require non-linear MIMO processing and are not considered in this paper).
The overall scheduling outcome can be represented by the matrix $\mathbf{X} = \left[\mathbf{x}_{1}~\mathbf{x}_{2}~...~\mathbf{x}_{B} \right]$ where $\mathbf{x}_{b}$ is the vector of UE indices selected for the $b$-th RBG.

Selecting which UEs to schedule jointly is inherently a combinatorial problem.
As for each RBG, the scheduler must choose a subset of $K_{b} \leq M$ users out of $U$ candidates, the total number of unique subsets (excluding permutations) is given by:

\begin{equation}
    \sum_{j=1}^{K_{b}}\left(\begin{array}{c}U\\ j\end{array}\right) = \sum_{j=1}^{K_{b}} \frac{U!}{j!\left(U-j \right)!} 
    \label{eq:user-selection}
\end{equation}

The search space, therefore, grows substantially with the number of candidate users $U$ and the number of simultaneously multiplexed UEs $K_{b} \leq M$.
Even moderate system configurations quickly lead to a prohibitively large number of combinations, making exhaustive search infeasible within scheduling timescales.
This combinatorial growth underscores the need for low-complexity yet effective user selection algorithms, such as the one proposed in the following section.

\vspace{-8pt}

\section{The Space Splitting-based User Selection Method} 
\label{sec:user-selection}

In this section, we present the proposed Space Splitting-based User Selection (SS-US) algorithm, a novel approach for user selection enabling scalable multi-user scheduling in existing and emerging delay-critical, massive-connectivity scenarios while preserving low complexity and low latency.
Three core principles guide the design of SS-US.
First, SS-US replaces combinatorial exploration of user subsets by partitioning the spatial degrees of freedom into orthonormal subspaces.
Second, it avoids iterative dependencies between selected users by relying on one-shot correlation evaluation per spatial direction.
Third, it enables intrinsic parallelism by constructing multiple independent spatial hypotheses that can be evaluated concurrently.
In contrast to SUS~\cite{sus} and GZF~\cite{gzf}, which rely on sequential user addition and repeated orthogonality checks, projections, or matrix inversions, SS-US evaluates user compatibility only once per spatial direction.
Compared to mCore+~\cite{mcore-sched}, which incurs exponential complexity due to exhaustive user selection, SS-US maintains bounded and tunable complexity even for large MIMO dimensions.
These principles directly underpin the scalability and low-complexity behavior shown in Sec.~\ref{sec:evaluation}.

\begin{figure}
    \centering
       \includegraphics[width=0.62\columnwidth]{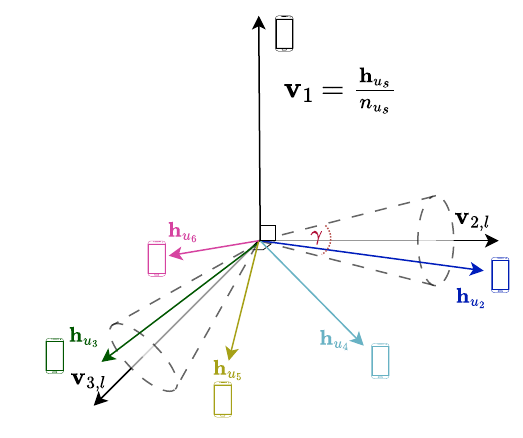}
\caption{Representing the selection of users based on their channel vectors $\mathbf{h}_{u_2}$  and $\mathbf{h}_{u_3}$ that are closely aligned with vectors $\mathbf{v}_{2,l}$ and $\mathbf{v}_{3,l}$ of the orthonormal base $\mathbf{V}_{l}$. Here, $n_{u_s}=\Vert \mathbf{h}_{u_s} \Vert$.}
    \label{fig:userSel}
    \vspace{-15pt}
\end{figure}

The algorithm starts by selecting the user whose channel vector $\mathbf{h}_{u}$ exhibits the largest single-stream rate $r_{u}$: 

\begin{equation}
    r_{u} =  \log_{2}{\left(1+\frac{\mathbf{h}_{u}^H  \mathbf{h}_{u} }{N_0}\right)}
    \label{eq:rate}
\end{equation}

\noindent where $N_{0}$ is the noise power and \( (\cdot)^H \) denotes the Hermitian transpose.
The channel vector of the strongest user, denoted \( \mathbf{h}_s \), is then normalized and used as the common first basis vector, $\mathbf{v}_{1} = \mathbf{h}_{s}/\lVert \mathbf{h}_{s} \rVert$ for all $L$ parallel orthonormal bases generated.
Starting from this shared vector $\mathbf{v}_{1}$, the algorithm constructs $L$ independent $M$-dimensional orthonormal bases $\mathbf{V}_{l} = \left[\mathbf{v}_{1},\mathbf{v}_{2,l},...,\mathbf{v}_{M,l} \right]$, where $M$ is the number of BS antennas.
For each basis, the remaining $M-1$ vectors are drawn from random independent vectors and orthogonalized using the Gram-Schmidt process. 
By generating $L$ independent orthonormal bases from the same strongest user, SS-US expands the spatial search for highly orthogonal users (see Fig.~\ref{fig:userSel}) while also enabling parallel processing to reduce runtime latency.

To evaluate the compatibility between a user channel $\mathbf{h}_{u}$ and a basis vector $\mathbf{v}_{m,l}$ of the $l$-th orthonormal basis, SS-US employs a well-known correlation metric:


\begin{equation}
    \Delta(\mathbf{h}_{u}, \mathbf{v}_{m,l}) = \frac{\lvert \mathbf{h}_{u}^H \mathbf{v}_{m,l}\rvert}{ \lVert \mathbf{h}_{u} \rVert \lVert \mathbf{v}_{m,l} \rVert }\ 
    \label{eq_corrcoef}
\end{equation}

Each basis vector $\mathbf{v}_{m,l}$ defines a ``directional cone'' characterized by a vertex angle $\gamma$ within which candidate users are evaluated, as conceptually illustrated in Fig.~\ref{fig:userSel}.
A user is accepted for a given basis direction only if the correlation value is above a minimum threshold $\alpha = \cos\left(\gamma/2\right)$ (see Fig.~\ref{fig:userSel}).
Fundamentally, increasing the values of $\alpha$ enforces the required orthogonality between users, but restricts the candidate pool.
This can result in having only low-norm users to be selected for an orthonormal basis, which may degrade the achieved spectral efficiency.
The optimal values of $\alpha$ are determined experimentally, as demonstrated in Sec.~\ref{subsec:eval-spectral} (note that experimental selection of conceptually similar parameters is common in related approaches, see e.g.~\cite{sus}).
The parameter $\alpha$ is also responsible for determining the resulting number of selected users $K \leq M$, in case no user candidate exceeds the threshold for an orthonormal basis vector.

The algorithm selects, for each basis vector $m$ of each basis~$l$, the user that maximizes the product $w_{u,m,l}=\Delta(\mathbf{h}_{u}, \mathbf{v}_{m,l}) \cdot r_{u}$. 
This combined metric favors users who exhibit both high channel gain and near orthogonality to previously selected users, thereby balancing signal strength and spatial compatibility.

SS-US naturally exploits multiple degrees of parallelism, making it well-suited for high-density and latency-sensitive connectivity scenarios.
Multiple orthonormal bases can be instantiated in parallel from the same initial user, each exploring alternative orthogonal subspaces; increasing the number of bases improves the likelihood of identifying highly orthogonal user sets at the cost of additional computation.
Within each basis, the correlation metric $\Delta(\mathbf{h}_{u}, \mathbf{v}_{m,l})$ and its rate-weighted evaluation can be computed independently for each spatial direction, enabling parallel execution across cores or hardware threads.
Furthermore, the candidate user set can be partitioned to evaluate correlations concurrently, further reducing latency.


The Algorithm~\ref{alg_osas} outlines the main steps of SS-US.
First, the channel vector with the largest norm is selected in Step~5. 
Then, $L$ orthonormal bases $\mathbf{V}_{l}$ are generated, starting from $\mathbf{v}_1$ being initialized to the normalized channel vector with the largest norm.
Next, the product between the correlation metric and the achievable user rate is computed in Step 9.
Users who maximize the product, subject to the threshold condition, are selected and removed from the candidate list (Steps 11-12).
Finally, the algorithm selects the candidate user set with the highest average channel norm among its selected users, maximizing spectral efficiency (Step~19).

\begin{algorithm}[t]
\caption{Space Splitting-based User Selection (SS-US)}\label{alg_osas}
\begin{algorithmic}[1]
\State \textbf{procedure} $\mathbf{x}_b=$ \texttt{SS-US} $\left( \mathbf{H}_{b} \right)$ \Comment{User selection}
\newline
\textbf{Input:}  $\mathbf{H}_b$: $(M \times \lvert \mathcal{U} \rvert )$ matrix of channel coefficients\newline
\textbf{Output:} $\mathbf{x}_b$: $({K_{b}} \times 1 )$ vector  of indices of UEs allocations
\For{$u=1$ to $\lvert \mathcal{U} \rvert$} \Comment{find column with largest norm}
    \State $n_u = \lVert \mathbf{H}_b \left( :,u \right)  \rVert_2$ \Comment{norm of $u$-th column}
   
\EndFor
  \State [$n_{u_s}$] = $\mathrm{max} \left(n_u \right)$ 
   \State Generate $L$ orthonormal bases $\mathbf{V}_{l}$ starting from $ \mathbf{v}_1=\mathbf{H}_b \left( :,u_s \right)/n_{u_s}$ via the Gram-Schmidt process.
\For{$l=1$ to $L$}
\For{$k=2$ to ${K} \leq M$}
    \State $[\mathbf{x}_b\left( k \right)] = \arg\max_{u \in [1,U]} \left( \Delta(\mathbf{h}_u, \mathbf{v}_{k,l}) \cdot r_u \right)$ \Comment{Find index of user maximizing the product of correlation (Eq.~\ref{eq_corrcoef}) and achievable rate}

\If{$\Delta(\mathbf{h}_{u^{*}}), \mathbf{v}_{k,l}) \ge \alpha)$}
        \State $\mathbf{x}_b(k) \leftarrow u^{*}$ \Comment{accept user for direction $\mathbf{v}_{k,l}$}
         \State $\mathbf{H} \left( :,\mathbf{x}_b \right) = []$ \Comment{remove selected user from candidates}
    \Else
        \State \textbf{continue} \Comment{skip if $\alpha$ threshold is not met}
    \EndIf
\EndFor
\State Calculate the mean metric  of selected users for matrix $l$ (i.e., $\hat w_{u^{*},l} =\frac{1}{K_b}\sum_{m=1}^Mw_{u^{*},m,l}$)
\EndFor
\State Select matrix with the highest mean metric (i.e., $\hat w_{u^{*},l}$)
\end{algorithmic}
\end{algorithm}

\vspace{-5pt}
\section{Evaluation}
\label{sec:evaluation}

\subsection{Experimental Setup}
\label{subsec:eval-exp-setup}

The proposed user selection method is evaluated using a MATLAB\textsuperscript{\textregistered} link-level simulator configured with the 3GPP Clustered Delay Line (CDL)-B channel model~\cite{3gppTR38901}, which characterizes an urban environment with rich scattering and predominantly non-line-of-sight (NLOS) propagation conditions.
User mobility is set to 3~km/h, representing pedestrian speeds. 


A hexagonal cell layout is populated with active single-antenna UEs uniformly distributed across the coverage area of the analyzed cell of $M$ antennas.
The total number of candidate users varies according to predefined pools $U$ of 20, 50, and 100 UEs, enabling assessment under different user densities.
In each simulation, $K$ users are selected for transmission in the linear MIMO system with 4 or 8 antennas, hence $K\leq M$.
Full path loss compensation is assumed, ensuring that each transmission achieves the target received power $P_0$ at the BS according to the CDL-B propagation conditions.
Consequently, the resulting signal-to-noise ratio (SNR) values are derived from the channel realizations without imposing maximum transmit power limits.

The modulation and coding schemes (MCSs) follow the standard 5G NR MCS Index Table~1~\cite{3gppTS38214}, selected to maximize achievable data rates under given channel conditions, supporting up to 64-QAM with a target block error rate (BLER) of 10\%.
Channel state information (CSI) is assumed to be perfect at the receiver. 
User selection decisions are based on the average channel quality over two RBGs.
The results are obtained by averaging 1000 simulation executions for each scenario.
The key simulation parameters are summarized in Table~\ref{tab:simu-param}.
This setup enables consistent evaluation across diverse spatial multiplexing levels and user densities.

\begin{table}[]
    \centering
    \caption{Main simulation parameters.}
    \begin{tabular}{|c|c|}
     \hline
     \textbf{Parameter} & \textbf{Value} \\
     \hline
      Noise figure & 5~dB \\
      \hline
      Bandwidth & 20~MHz \\
      \hline
      Carrier frequency & 3.5~GHz \\
      \hline
      Shadowing std. dev. ($\sigma$) & 4~dB \\
      \hline
      Path loss compensation & 1 \\
      \hline
      Target receiv. power ($P_{0}$) & \{-90, -95\}~dBm\\
      \hline
      Inter-site distance & 500~m \\
      \hline
      Num. of candidate users & \{20, 50, 100\} \\
      \hline
      Num. of BS antennas & \{4, 8\} \\
      \hline
      
    \end{tabular}
    \label{tab:simu-param}
    \vspace{-15pt}
\end{table}

\subsection{Complexity Analysis}
\label{subsec:eval-complexity}

Computational complexity is critical for scalability in dense user and large-array regimes, where user selection must operate at microsecond timescales. This subsection compares the asymptotic complexity of SUS~\cite{sus}, Greedy Zero-Forcing (GZF)~\cite{gzf}, and mCore+~\cite{mcore-sched} with the proposed algorithm.


The dominant cost components of user selection arise from:
i) inner product or projection operations between user channel vectors, with complexity on the order of $\mathcal{O}(M)$;
ii) matrix operations such as QR updates or inversions, with complexity $\mathcal{O}(MK^{2})$ or $\mathcal{O}(K^{3})$, and
iii) iterations over all candidates, with a multiplicative factor $U$, or exhaustive searches.
Ignoring lower-order operations, such as sorting or scalar divisions, the overall complexity for each method can be derived as follows:

\paragraph{SUS} This method iteratively enforces near-orthogonality among selected users by projecting candidate channels onto the orthogonal complement of the selected set~\cite{sus}.
The first user is chosen as the one with the largest channel norm, requiring $\mathcal{O}(UM)$ operations.
In each subsequent iteration, projections and residual updates are computed for all remaining candidates, with $\mathcal{O}(UM)$ cost per iteration, repeated until $K$ users are selected.
With $K-1$ iterations remaining, the overall complexity is therefore:
\vspace{-5pt}
\begin{equation}
    C_{SUS} = \mathcal{O}(UM + \sum_{k=1}^{K-1} (UMk)) \approx \mathcal{O}(UM(1+K^2)
\end{equation}

\paragraph{GZF} This algorithm iteratively selects users by maximizing the incremental performance gain (e.g., sum-rate) achieved when adding a candidate to the current set.
Starting from the strongest user, each iteration $k$ evaluates all remaining candidates by constructing and inverting the Gram matrix $\mathbf{G}$, which incurs a cost of $\mathcal{O}(Mk^2)$ and $\mathcal{O}(k^3)$, respectively.
As this process repeats for roughly $U$ candidates across $K-1$ iterations, the total complexity can be approximated as:
\vspace{-5pt}
\begin{equation}
\begin{aligned}
    C_{GZF} = UM + \sum_{k=1}^{K-1}(U-k)(Mk^2+k^3) \\ C_{GZF} \approx \mathcal{O}(U(M+ MK^{3}+K^{4})) 
\end{aligned}
\end{equation}


The strong dependence on $K^4$ makes GZF computationally demanding and poorly scalable to large user selections and pools, despite typically yielding better performance than SUS.




\paragraph{mCore+} This method performs user selection based on chordal distance computations~\cite{mcore-sched}.
First, the method reduces the candidate pool from $U$ to $2M$ by selecting strongest users, at cost $\mathcal{O}(UM)$.
Within this reduced set, chordal distances are computed ($\mathcal{O}(M)$ per comparison) to identify $M$ quasi-orthogonal users.
Finally, an exhaustive search over the remaining $M$ users determines the subset that maximizes the sum rate, yielding the total complexity:
\vspace{-5pt}
\begin{equation}
\begin{aligned}
    C_{mCore+} = \mathcal{O} \left( UM + M^3 + \sum_{k=1}^M\left(\begin{array}{c}M\\ k\end{array}\right)(Mk^2+k^3) \right)
\end{aligned}
\end{equation}

Although mCore+ provides high spectral efficiency, its complexity grows rapidly with the number of MIMO layers, making the exhaustive step prohibitive beyond four streams.
Furthermore, as originally designed for downlink, it expects to receive SVD information from UEs, while in uplink, this computation must be performed at the BS, further increasing complexity.

\paragraph{SS-US} The proposed SS-US method begins by selecting the strongest candidate (highest channel norm) as the seed, with cost $\mathcal{O}(UM)$. 
From this seed, the algorithm constructs up to $L$ orthonormal basis sets, each spanning a candidate subspace. The parameter $L$ provides a tunable trade-off between computational cost (via parallel basis processing) and spatial exploration diversity.
For each basis, an $M$-dimensional correlation analysis is performed between the users' channels and the basis directions, identifying the one that best aligns with each spatial dimension.
Generating one basis from the initial user costs $\mathcal{O}(M^{3})$ while evaluating correlations for all $U$ users across $M-1$ dimensions requires $\mathcal{O}(UM(M-1))$.
Hence, the total complexity is expressed as:

\begin{equation}
    C_{SS-US} = \mathcal{O} (UM + L \left( M^{3} + UM(M-1) \right)
\end{equation}

Since the computations for the $L$ bases are embarrassingly parallel, the effective runtime can approach that of a single basis evaluation under concurrent execution.
By tuning $L$, SS-US enables a controllable complexity-performance trade-off between lightweight operation and broader spatial diversity.
Although its formal asymptotic complexity resembles that of SUS ($C_{SUS} \approx \mathcal{O}\left( UM + UM^3\right)$ when $K \rightarrow M$), SS-US is considerably more efficient.
Unlike SUS, which sequentially computes costly null-space projections for remaining users, SS-US only analyzes lightweight correlations per basis vector, enabling scalable, parallelizable operation and competitive spectral efficiency, well aligned with the low-complexity and scalability demands of emerging massive, delay-critical connectivity scenarios.

\begin{figure}
    \centering
    \includegraphics[width=0.95\columnwidth]{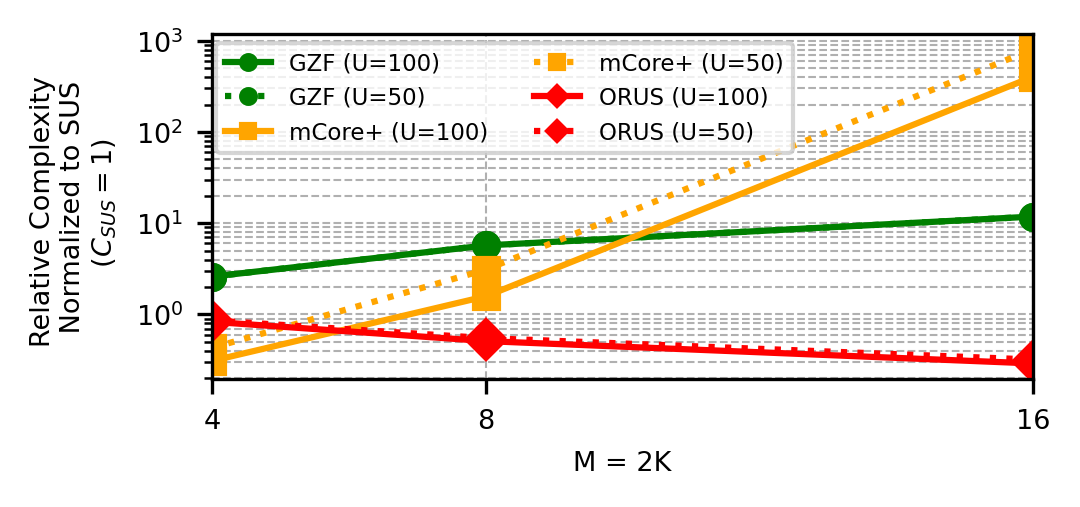}
    \caption{Relative complexity normalized to SUS ($C_{SUS}=1$) in a linear MU-MIMO system ($M$ antennas $= 2K$ users).}
    \label{fig:complexity}
    \vspace{-15pt}
\end{figure}

Fig.~\ref{fig:complexity} compares the relative computational cost of the evaluated schemes, normalized to the SUS baseline ($C_{\text{SUS}}=1$), in a linear-processing MU-MIMO system with twice as many antennas as users ($M=2K$).
SS-US achieves a significant reduction in complexity compared to SUS and the other baselines, particularly for configurations with more than four antennas.
For instance, with $M=2K=8$, SS-US requires nearly half the computational cost of SUS and over 10$\times$ lower complexity than GZF.
Compared to mCore+, SS-US also demonstrates a substantial advantage that becomes more pronounced for $M>4$, achieving over three orders of magnitude (over 2500$\times$) lower complexity in larger configurations, with high parallelization opportunities.

\subsection{Spectral Efficiency Analysis}
\label{subsec:eval-spectral}

\begin{figure}
    \centering
    \includegraphics[width=0.999\columnwidth]{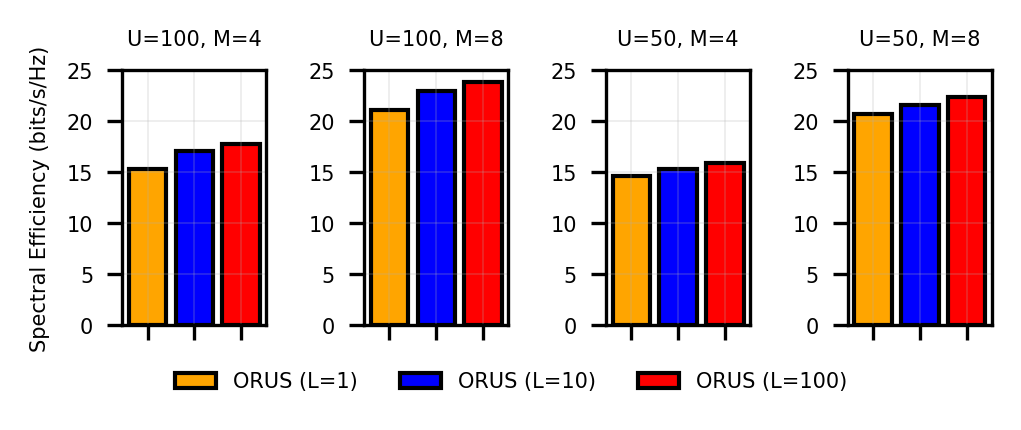}
    \caption{Increasing the number of parallel bases $L$ yields higher spectral efficiency at the cost of a linear increase in complexity.}
    \label{fig:matrix-gen}
    \vspace{-15pt}
\end{figure}

We now assess the spectral efficiency of SS-US relative to the considered baselines, including random selection as a lower performance bound.
We first analyze the impact of varying the number $L$ of parallel random initializations for the orthonormal matrix, with $L=\{1,10,100\}$ (see Fig.~\ref{fig:matrix-gen}).
Increasing $L$ expands the orthogonality search space by generating multiple parallel candidate matrices, but also raises computational complexity.
As shown in Fig.~\ref{fig:matrix-gen}, higher $L$ values improve spectral efficiency, with gains exceeding 15\% for specific cases.
However, the complexity rises substantially, and based on the analysis from Sec.~\ref{subsec:eval-complexity}, it increases linearly with $L$.


\begin{figure}
    \centering
    \includegraphics[width=0.95\columnwidth]{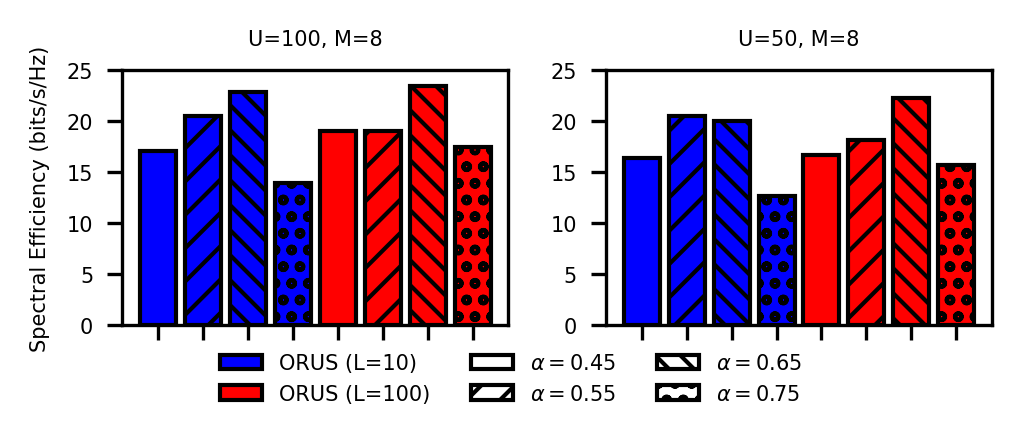}
    \caption{SS-US performance for indicative $\alpha$ values: overly strict threshold can limit user pool and lower spectral efficiency.}
    \label{fig:alpha-compare}
    \vspace{-15pt}
\end{figure}

As described in Sec.~\ref{sec:user-selection}, the SS-US algorithm also includes a tunable correlation threshold $\alpha$, which defines the minimum orthogonality required between candidate users and basis vectors.
Increasing $\alpha$ enforces a stricter orthogonality criterion.
The impact of $\alpha$ on spectral efficiency is illustrated in Fig.~\ref{fig:alpha-compare}.
Higher $\alpha$ values promote a more orthogonal user set, but this constraint can exclude users with strong channel norms, reducing overall spectral efficiency.
Conversely, smaller $\alpha$ values relax the orthogonality condition, potentially admitting higher-norm users, at the cost of increased intra-user interference.
When multiple orthonormal bases are generated (e.g., $L=10$), higher $\alpha$ values tend to improve performance for sufficiently large user pools.
As shown in Fig.~\ref{fig:alpha-compare}, for $U=100$, increasing $\alpha$ from 0.45 to 0.65 yields nearly a 35\% gain for \textit{SS-US (L=10)}.
This occurs because the presence of several parallel bases already enhances the search space for orthogonal users, allowing the algorithm to maintain both high orthogonality and high user norms.

\begin{figure}
    \centering
    \includegraphics[width=0.999\columnwidth]{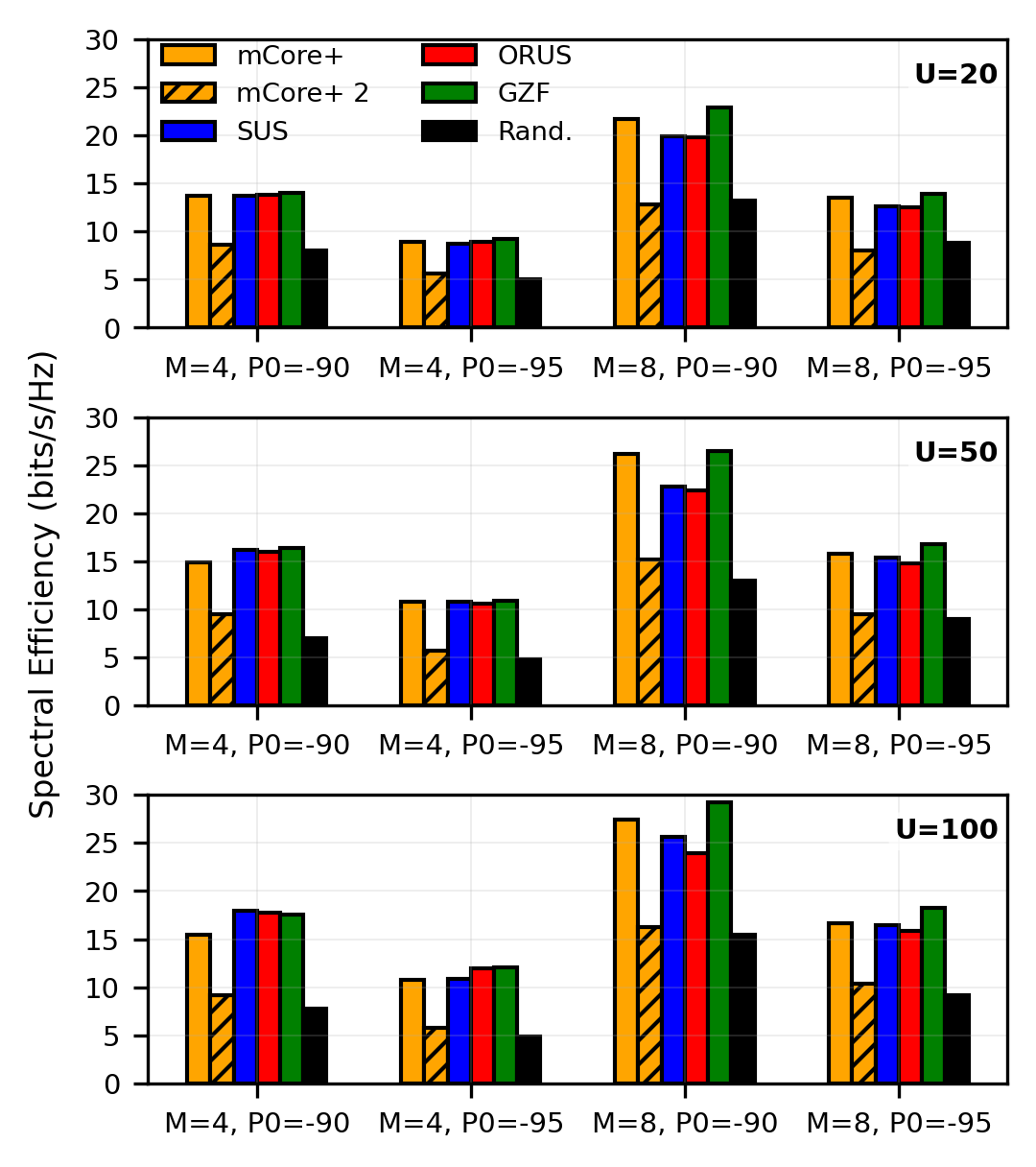}
    \caption{Comparison of SS-US against baselines~\cite{sus,mcore-sched,gzf}.}
    \label{fig:major-compare}
    \vspace{-25pt}
\end{figure}

Finally, Fig.~\ref{fig:major-compare} compares SS-US against the baseline user selection algorithms.
Across most evaluated scenarios, SS-US achieves spectral efficiency values comparable to the best-performing methods, despite being considerably less computationally demanding.
The observed differences typically remain within a $\pm$~3-8\% range and are often statistically insignificant according to $t$-tests.
For cases where $M=4$, SS-US performs on par with all the baselines considering all the user pool sizes, having comparable complexity to mCore+ and SUS and less than half of the complexity of GZF (see Fig.~\ref{fig:complexity}).
In low-SNR regimes ($P_{0}=-95~dBm$), when $M=8$, for all the user pool sizes, the spectral efficiency of SS-US can also approximate all the other baselines, with maximum relative gaps close to 10\%.
For the high-SNR regimes ($P_{0}=-90~dBm$), when $M=8$, SS-US still performs similarly to the baselines, and the gaps on spectral efficiency widen as the user pool size increases, particularly for GZF and mCore+.
However, for $M=8$, the complexity of SS-US is half of SUS, 3$\times$ less than mCore+, and over 20 times less than GZF.
The simplified variant \textit{mCore+~2}, introduced to constrain its search space and complexity with a similar $\alpha$ threshold, exhibits a clear performance loss, confirming that excessive pruning harms selection diversity.
Another consistent trend is the general degradation of spectral efficiency when user diversity and channel conditions worsen (i.e., smaller $U$ or lower $P_{0}$), where all algorithms approach the performance of random selection.
Nonetheless, SS-US remains consistently close to the top-performing methods across all MIMO configurations.
These results validate SS-US as a scalable, low-complexity alternative for next-generation massive MU-MIMO, maintaining high spectral efficiency while supporting large user pools and MIMO layers under delay-critical traffic.

\vspace{-2pt}

\vspace{-5pt}
\section{Conclusions}
\label{sec:conclusions}

\vspace{-3pt}
The rise of AI-native traffic, characterized by increasingly uplink-oriented and delay-sensitive communication patterns, amplifies the need for scalable MU-MIMO solutions capable of supporting frequent scheduling and user selection under massive connectivity.
While existing schemes provide valuable heuristics, their scalability degrades sharply as the number of users and spatial layers grows.
In this work, we introduced the Space Splitting-based User Selection (SS-US) algorithm, which constructs orthonormal bases and applies correlation-driven selection to identify spatially compatible users with minimal computational cost.
Our results show that SS-US can offer over three orders of magnitude lower complexity and massive parallelization opportunities while achieving comparable spectral efficiency to practical, high-performing baselines.
These attributes make SS-US well-suited for next-generation MU-MIMO deployments facing high user densities and AI-native traffic dynamics.
Future work will focus on investigating new orthonormal matrix generation strategies, supporting single-user MIMO, and integrating SS-US into practical system schedulers that jointly optimize time, frequency, spatial layers, and power usage, while incorporating implementation awareness to exploit hardware-software flexibility.

\vspace{-4pt}
\bibliographystyle{IEEEtran}
\bibliography{IEEEabrv,refs}

\end{document}